\documentclass{elsart} 
\usepackage{amssymb} 

\begin{document}

\begin{frontmatter}  

\title{A possible deformed algebra and calculus 
inspired in nonextensive thermostatistics}
\author{Ernesto P. Borges}
\ead{ernesto@ufba.br}
\address{Escola Polit\'ecnica, Universidade Federal da Bahia \\
Rua Aristides Novis 2, 40210-630 Salvador-BA, Brazil 
\\ \and \\
Centro Brasileiro de Pesquisas F\'\i sicas \\
R. Dr. Xavier Sigaud, 150, Urca, 22290-180 Rio de Janeiro-RJ, Brazil
} 

\begin{abstract} 
We present a deformed algebra related to the $q$-exponential and the 
$q$-logarithm functions that emerge from nonextensive statistical mechanics. 
We also develop a $q$-derivative (and consistently a $q$-integral) 
for which the $q$-exponential is an eigenfunction.
The $q$-derivative and the $q$-integral have a dual nature, 
that is also presented.
\end{abstract}  


\end{frontmatter}  

\date{\today} 
\maketitle   

\section{Introduction}

The definition of the so-called nonextensive entropy,
$
S_q\equiv k(\sum_i^W p_i^q -1)/(1-q)
$
($q\in {\mathbb{R}}$ is the nonextensive entropic index, 
$k$ is a positive constant which defines the unity in which $S_q$ is measured), 
and its applications to nonextensive thermostatistics (NEXT) 
\cite{ct:1988,ct-evaldo} have opened new possibilities for the 
thermodynamical treatment of complex systems \cite{a.cho:science}, 
with a continuously increasing number of works 
(for an updated bibliography, see Ref. \cite{http}).
Nonextensive statistical mechanics has been treated basically along three
(complementary) lines: formal mathematical developments 
\cite{sumiyoshi-raja:science,plastino:science}
(in which this paper is included), 
observation of consistent concordance with experimental (or natural) behavior
\cite{sumiyoshi-raja:science},
and theoretical physical developments \cite{vito:science} 
(see Ref. \cite{ct:opening-cagliari2001} for some epistemological remarks 
on NEXT).

The basis of the formal developments inspired in NEXT is
the definition of deformed expressions for the beautiful
and ubiquitous logarithm and exponential functions, namely,
the $q$-logarithm and the $q$-exponential, first proposed within this
context in 1994 \cite{ct:quimicanova}:
\vspace{-0.5cm}
\begin{eqnarray}
 \label{qlog-qexp}
 \ln_{q} x \equiv \frac {x^{1-q}-1}{1-q} \quad (x>0),
 \qquad
 e_q(x) \equiv 
   [1+(1-q)x]^{\frac{1}{1-q}}_+
 \quad (x,q \in {\mathbb{R}}),
\end{eqnarray}
where $[A]_+\equiv\mbox{max}\{A,0\}$.
Note that $e_q(x)=+\infty$ if $q>1$ and $1+(1-q)x\le 0$.
Of course there are infinitely many ways of generalizing a function,
and many have already been proposed, for instance (see Ref. \cite{kassel}),
$\exp_q x \equiv \sum_{n=0}^\infty x^n/[n]_q!$,
with $[n]_{q}!=\prod_{j=1}^{n} [j]_{q}$
and $[j]_{q}=(q^{j}-1)/(q-1)$ and also $[0]_{q}!=1$,
but we focus here specifically on those ones that naturally emerge from NEXT
(Eq.~(\ref{qlog-qexp})).
Some works have appeared developing properties of such $q$-deformed
functions, e.g., Ref.
\cite{epb:jpa1998,ct:bjp1999,epb:jpa1999,ct:springer,yamano,naudts},
and have also stimulated the definition of other (alternative) deformations,
as in Ref. \cite{giorgio:2001,giorgio:next2001,giorgio:relativity}.

It is remarkable that in less than one decade (since Ref. \cite{ct:quimicanova})
these functions ($\ln_q x$ and $e_q(x)$) have become so popular among the 
practitioners of NEXT that in many papers they are referred to as if they were 
known since long, which is a very positive symptomatic behavior.
And they are known since long, indeed: 
the $q$-exponential is a particular solution of Bernoulli's equation, known
by Leibniz (according to Refs.~\cite{BoyceDiPrima,Kreyszig}).
(The independent re-discovering of laws, natural phenomena etc. is much more 
frequent in science than one outsider would be inclined to think.)

This paper presents some formal, and curious, properties of $q$-exponentials 
and $q$-logarithms, hopefully inspiring for further developments.
In some sense, we follow the work of Kaniadakis, who has explored his
own $\kappa$-deformation of the exponential and logarithm functions 
\cite{giorgio:2001,giorgio:next2001,giorgio:relativity}.

\section{$q$-Algebra%
\protect\footnote{We point out that parts of the contents of this section have 
appeared independently and almost simultaneously in the early versions of 
Ref.~\protect\cite{alexandre} and also in the early version of the present 
work, as explained in the Acknowledgments.}}
\label{sec:q-algebra}

We focus on the (now well known) following relations:
\vspace{-0.5cm}
\begin{eqnarray}
 \label{ln_q(xy)}
 \ln_q(xy)=\ln_q x + \ln_q y + (1-q)\ln_qx\ln_qy
\end{eqnarray}
(which justifies the terminology ``nonextensive'')
and
\vspace{-0.5cm}
\begin{eqnarray}
 \label{e_q^xe_q^y}
 e_q(x)\,e_q(y)=e_q(x+y+(1-q)xy),
\end{eqnarray}
which holds if $e_q(x)$ and $e_q(y)$ differ from zero and from $+\infty$.
These properties inspire us to define a generalization of the sum operation
between two numbers $x$ and $y$:
\vspace{-0.5cm}
\begin{eqnarray}
  \label{q-sum}
  x \oplus_q y \equiv x + y + (1-q)xy,
\end{eqnarray}
which brings the usual sum as a particular case $\oplus_1 \equiv+$.
The $q$-sum is commutative
($x\oplus_q y=y \oplus_q x$),
associative
($x \oplus_q (y \oplus_q z) = (x \oplus_q y) \oplus_q z$), 
but it is not distributive in relation to the usual multiplication
($a(x \oplus_q y)\ne (ax\oplus_q ay)$).
The neutral element of the $q$-sum is zero,
$x \oplus_q 0 =  x$.
We can define the opposite (or inverse additive) of $x$ (calling it 
$\ominus_q x$)
as the element that, when $q$-summed with $x$, yields the neutral element:
$x \oplus_q (\ominus_q x) =0$.
So, we have
\vspace{-0.5cm}
\begin{eqnarray}
  \label{q-minus}
  \ominus_q x \equiv \frac{-x}{1+(1-q)x} \qquad (x\ne1/(q-1)).
\end{eqnarray}
This definition permits us to define the $q$-difference, as the $q$-sum
with the $q$-opposite
\vspace{-0.5cm}
\begin{eqnarray}
  \label{q-diff}
  x \ominus_q y \equiv x \oplus_q (\ominus_q y)
          =      \frac{x-y}{1+(1-q)y} \qquad (y\ne1/(q-1)).
\end{eqnarray}
The $q$-difference obeys
$x\ominus_q y =\ominus_q y\oplus_qx$ and
$x \ominus_q (y \ominus_q z) = (x \ominus_q y) \oplus_q z$,
but
$  a(x\ominus_q y)\ne (ax\ominus_q ay)$.

We now search for a generalization of the multiplication operation 
in such a way that we can compactly rewrite Eqs.~(\ref{ln_q(xy)}) and 
(\ref{e_q^xe_q^y}) as
$
\ln_q(x \otimes_q y) = \ln_q x + \ln_q y
$
and
$
e_q(x) \otimes_q e_q(y) = e_q(x+y).
$
This leads us to the definition of the $q$-product between two numbers
\vspace{-0.5cm}
\begin{eqnarray}
  \label{q-product}
  x \otimes_q y \equiv \left[x^{1-q}+y^{1-q}-1\right]^{\frac{1}{1-q}}_+
  \qquad (x,y> 0).
\end{eqnarray}
The $q$-product is commutative ($x\otimes_q y=y\otimes_q x$) and associative
($x\otimes_q (y\otimes_q z)=(x\otimes_q y)\otimes_q z$),
provided $x\otimes_q y$ and $y\otimes_q z$ differ from zero and from infinity.
It is easy to see that the number one is the neutral element of the
$q$-product
($x \otimes_q 1 = x$),
and this permits us to define the inverse multiplicative, $1\oslash_q x$,
by means of
$x \otimes_q (1\oslash_q x) \equiv 1$.
We find
\vspace{-0.5cm}
\begin{eqnarray}
  \label{q-inverse}
  1\oslash_q \,x \equiv \left[2-x^{1-q}\right]^{\frac{1}{1-q}}_+
  \qquad (x\ge 0).
\end{eqnarray}
The relation $1\oslash_q \,(1\oslash_q \,x)=1$
holds only if $0\le x^{1-q}\le 2$.
It is curious to note that $1\oslash_q 0$ does not diverge for $q<1$.
The $q$-ratio is thus defined by
\vspace{-0.5cm}
\begin{eqnarray}
  \label{q-ratio}
  x\oslash_q \,y \equiv \left[x^{1-q}-y^{1-q}+1\right]^{\frac{1}{1-q}}_+
   \qquad (x,y>0),
\end{eqnarray}
and satisfies
$x\oslash_q y=1\oslash_q (y\oslash_q x)$,
provided $x^{1-q}\le 1+y^{1-q}$,
and also 
$x\oslash_q(y\oslash_qz)=(x\oslash_qy)\otimes_qz=(x\otimes_qz)\oslash_qy$,
provided $z^{1-q}-1\le y^{1-q}\le x^{1-q}+1$.
These $q$-algebraic relations permit us to express the properties of the
$q$-logarithm and the $q$-exponential in a more compact form:
\vspace{-0.7cm}
\begin{eqnarray}
  \ln_q(xy)=\ln_qx \oplus_q \ln_qy
  &
  \qquad \qquad
  e_q(x)\;e_q(y)=e_q(x\oplus_qy)
  \label{lnq(xy)}
  \\
  \ln_q(x\otimes_qy)=\ln_qx+\ln_qy
  &
  \qquad \qquad
  e_q(x)\otimes_q e_q(y)=e_q(x+y)
  \label{qfactoring}
  \\
  \ln_q(x/y)=\ln_qx\ominus_q\ln_qy
  &
  \qquad \qquad
  e_q(x)/e_q(y)=e_q(x\ominus_qy)
  \\
  \ln_q(x\oslash_qy)=\ln_qx-\ln_qy
  &
  \qquad \qquad
  e_q(x)\oslash_q e_q(y)=e_q(x-y)
  \label{qexp(x-y)}
\end{eqnarray}
The equalities hold only under certain restrictions, as indicated in the
following table
\vspace{-0.7cm}
\begin{eqnarray}
 \label{restrictions}
 \begin{array}{ll}
 \hline 
 x>0,y>0              & \qquad \qquad x\ge_q 0\hbox{ or }y\ge_q 0\cr
 x^{1-q}+y^{1-q}\ge 1 & \qquad \qquad x\ge_q 0\hbox{ and }y\ge_q 0\cr
 x>0,y>0              & \qquad \qquad y>_q0\cr
 x^{1-q}+1\ge y^{1-q} & \qquad \qquad x\ge_q 0\hbox{ or }y\ge_q 0
 \\ \hline
 \end{array}
\end{eqnarray}
Here, the notation $x\ge_q 0$ means $1+(1-q)x\ge 0$.
Among these relations, we call attention to Eq.~(\ref{qfactoring}), 
that shows how to $q$-factorize the kinetic and potential parts of the 
Hamiltonian of a system (and further $q$-factorize each one of them 
individually).
It is straightforward to define a $q$-power,
\vspace{-0.5cm}
\begin{eqnarray}
  x^{\otimes_q^n}\equiv
  \underbrace{x\otimes_qx\otimes_qx\otimes_q \cdots \otimes_qx}_{\mbox{n times}}
                 = \left[nx^{1-q}-(n-1)\right]^{\frac{1}{1-q}}_+,
\end{eqnarray}
with $x^{\otimes_1^n}\equiv x^n$.
It follows also straightforwardly the $q$-sum of $n$ identical terms
(see Eq.~(7) of Ref.~\cite{vives:prl}):
\vspace{-0.5cm}
\begin{eqnarray}
 \underbrace{x\oplus_qx\oplus_qx\oplus_q \cdots \oplus_qx}_{\mbox{n terms}}
                      =\frac{1}{1-q}\left\{\left[1+(1-q)x\right]^n-1 \right\}.
\end{eqnarray}
We call attention that this Equation may be the basis for the definition
of a different, non-commutative, $q$-product (that we could note $\odot_q$):
$n\odot_qx \ne x\odot_qn$, with $1\odot_qx=x$ and $n\odot_1x=nx$. 
There is another possibility of generalizing the $q$-algebra, along 
nonextensive lines, recovering the distributivity, as done in 
Ref. \cite{alexandre}, following Ref. \cite{giorgio:relativity}.
We don't explore these possibilities in the present work.

\section{$q$-Calculus}

We now focus on a possible $q$-calculus associated with NEXT.
Abe \cite{sumiyoshi:entropy} has shown that the Jackson's $q$-derivative,
${\mathcal{D}}_q f(x) \equiv [f(qx)-f(x)]/[(q-1)x]$,
(that is related to dilations) is intimately connected to nonextensive entropy, 
as the usual (Newtonian) derivative (that is related to translations)
is connected to Boltzmann-Gibbs entropy. 
But this Jackson's operator is not the proper one we are searching here.
The usual exponential function is invariant in relation to derivation,
$dy/dx=y$, or, in other words, $e^x$ is eigenfunction of the derivative
operator. The $q$-exponential satisfies $dy/dx=y^q$. So we can ask: what is the
operator for which the $q$-exponential is an eigenfunction?
The answer to this question may be found by following the steps of Kaniadakis
\cite{giorgio:2001,giorgio:next2001,giorgio:relativity}.
Let us define the $q$-derivative as
\vspace{-0.8cm}
\begin{eqnarray}
  \label{D_q}
    \displaystyle D_{(q)} f(x) \equiv 
                        \displaystyle\lim_{y\to x}\frac{f(x)-f(y)}{x\ominus_qy}
                           = 
			   \displaystyle [1+(1-q)x] \frac{df(x)}{dx}.
\end{eqnarray}
It is promptly verified that the $q$-exponential is eigenfunction of $D_{(q)}$.
We have the corresponding $q$-integral (noted as $\int\nolimits_{(q)}$) 
given by
\vspace{-0.5cm}
\begin{eqnarray}
  \label{int_q}
  \int\nolimits_{(q)} f(x)\;d_qx = \int \frac{f(x)}{1+(1-q)x}\,dx,
\end{eqnarray}
where
\vspace{-0.8cm}
\begin{eqnarray}
  \label{d_qx}
  d_qx\equiv \lim_{y\to x} x\ominus_q y
      =       \frac{1}{1+(1-q)x}dx.
\end{eqnarray}
Naturally
$
  \int_{(q)} D_{(q)}f(x)\;d_qx=D_{(q)} \int_{(q)}f(x)\;d_qx=f(x).
$
Eqs.~(\ref{D_q})--(\ref{d_qx}) are restricted to $1+(1-q)x\not=0$.
There is a dual derivative operator $D^{(q)}$, associated with $D_{(q)}$,
defined by
\vspace{-0.5cm}
\begin{eqnarray}
  \label{D^q}
  \displaystyle
   D^{(q)} f(x) \equiv\displaystyle\lim_{y\to x} \frac{f(x)\ominus_qf(y)}{x-y}
             =      \displaystyle \frac{1}{1+(1-q)f(x)}\, \frac{df(x)}{dx},
\end{eqnarray}
and its corresponding dual $q$-integral
\vspace{-0.5cm}
\begin{eqnarray}
  \label{int^q}
  \int\nolimits^{(q)} f(x) dx \equiv \int [1+(1-q)f(x)] f(x)\;dx.
\end{eqnarray}
Also here 
$\int^{(q)} D^{(q)}f(x)\;dx=D^{(q)}\int^{(q)}f(x)\;dx=f(x)$.
Similarly to the restrictions of $D_{(q)}$ and $\int_{(q)}$,
Eqs.~(\ref{D^q}) and (\ref{int^q}) are valid if $1+(1-q)f(x)\not=0$.
We have
$  D^{(q)}\ln_qx=1/x$, while $D^{(1)} \ln_qx= \frac{d}{dx} \ln_qx = 1/x^q$.
The usual derivative and the usual integral are obviously auto-dual.
The dual $q$-derivatives are related by
\vspace{-0.5cm}
\begin{eqnarray}
 D^{(q)}f(x) = \frac{1}{[1+(1-q)x]\,[1+(1-q)f(x)]}\,D_{(q)}f(x).
\end{eqnarray}

The $q$-derivatives follow the chain rule in the forms:
\vspace{-0.5cm}
\begin{eqnarray}
 D_{(q)}[f(x)\,g(x)]= D_{(q)}[f(x)]\,g(x) + f(x)\,D_{(q)}[g(x)]
\end{eqnarray}
and
\vspace{-0.6cm}
\begin{eqnarray}
 \begin{array}{lll}
  D^{(q)}[f(x)\,g(x)]&=&\displaystyle\frac{1}{[1+(1-q)f(x)g(x)]} \\
                     &&\times\left\{[1+(1-q)f(x)]\,D^{(q)}[f(x)]\,g(x) \right.
                     \\
                     && +\left.[1+(1-q)g(x)]\,f(x)\,D^{(q)}\,[g(x)]\right\}.
 \end{array}
\end{eqnarray}

\section{Concluding Remarks}

The relations presented are compact and relatively simple, though 
there are many exceptions that have to be taken into account originated 
from the cutoff that appears in the $q$-deformed exponential function
(see, for instance, restrictions (\ref{restrictions})), 
and it is of course important that we are aware of them.
Despite these limitations, it is remarkable that
many phenomena seem to be fairly fitted by $q$-exponentials, 
or functions that belong to its family.
Examples range over physical, biological, economical and social systems
(all of them may be classified into the category of {\em complex systems}).
Besides the observation of fair description of phenomena, there are
increasingly many works on theoretical developments of nonextensive
ideas.
Just to cite an important and very recent example, let us mention the 
exact analytical results by Baldovin and Robledo 
\cite{fulvio-alberto:pesin}
about the $q$-Pesin identity for the dynamics at the edge of chaos of the
logistic map. 
For the interested reader, we give the address of the web page \cite{http}.
Of course all these elements (mathematical framework, experimental evidence,
theoretical developments) are not yet enough to be conclusive,
and many efforts have been made in order to clarify the domain of validity
of Boltzmann-Gibbs statistical mechanics, of nonextensive statistical mechanics,
or even others that might happen in nature.

\section*{Acknowledgments} 

We gratefully acknowledge Giorgio Kaniadakis for calling us attention to
the existence of the duality in the $q$-derivatives.
We also thank Q. Alexandre Wang (together with his coauthors) for sending to us 
their paper Ref.~\cite{alexandre}, while it was in its {\em e-print} form 
[{\tt http://arXiv.org/math-ph/0303061}], soon after he took notice of the 
early {\em e-print} version of the present work 
[{\tt http://arXiv.org/ cond-mat/0304545}].
These works present many similarities, specially regarding the $q$-algebra
(Section \ref{sec:q-algebra}). 
I am deeply indebted to Jan Naudts, who introduced many valuable improvements
on this work.
Last but not least, the unforgettable hospitality of the organizers of 
NEXT 2003 is warmly acknowledged.


\end{document}